\documentstyle[preprint,aps]{revtex}
\begin{document}
\def\brho{\mbox{\boldmath $\rho $}}
\draft
\renewcommand{\baselinestretch}{1.0}
\renewcommand{\theequation}{\arabic{equation}}
\input epsf

\title{\bf{Buckling, Fluctuations and Collapse in Semiflexible
           Polyelectrolytes}}
\author{ Per Lyngs Hansen $\dagger$$\ddagger$,
         Rudi Podgornik$\thanks{To whom
           correspondence should be addressed. E-mail address:
           rudi@helix.nih.gov}~\dagger$$\ddagger$$\S$ ,
         Daniel Sven\v sek $\S$, \\ and
         V. Adrian Parsegian$\dagger$}
\address{$\dagger$ Institute for Theoretical Physics, 
University of California, Santa Barbara, CA 93106-4030}

\address{$\ddagger$ Laboratory of Physical and Structural Biology, 
National Institute of Child Health and Human Development, 
National Institutes of Health, Bethesda, MD 20892-5626}

\address{$\S$ Department of Physics, Faculty of Mathematics and 
Physics, University of Ljubljana, SI-1000 Ljubljana, Slovenia}

\address{$\S$ Department of Theoretical Physics, J. Stefan Institute, SI-1000 Ljubljana, Slovenia}

\date{\today}
\maketitle
\begin{abstract}
We present a systematic statistical mechanical analysis of the 
conformational properties of a stiff polyelectrolyte chain with 
intrachain attractions that are due to counterion correlations.  We 
show that the mean-field solution corresponds to an Euler-like 
buckling instability.  The effect of the conformational fluctuations 
on the buckling instability is investigated, first, qualitatively, 
within the harmonic (``semiclassical'') theory, then, systematically, 
within a $1/d$-expansion, where $d$ denotes the dimension of embedding 
space.  Within the ``semiclassical'' approximation, we predict that 
the effect of fluctuations is to renormalize the effective persistence 
length to smaller values, but not to change the nature of the 
mean-field ({\sl i.e.}, buckling) behavior.  Based on the 
$1/d$-expansion we are, however, led to conclude that thermal 
fluctuations are responsible for a change of the buckling behavior 
which is turned into a polymer collapse.  
A phase diagram is constructed in which a sequence of collapse 
transitions terminates at a buckling instability that occurs at a 
place that varies with the magnitude of the bare persistence length of 
the polymer chain, as well as with the strength and range of the 
attractive potential.


\end{abstract}

\pacs{61.20.Qg, 61.25.Hq, 87.15.Da}
\section{Introduction}

Recently, the phenomenon of DNA condensation has been subject to 
renewed interest \cite{vic}, \cite{cosb}.  It appears that the attractive 
forces generated by the correlated fluctuations of \cite{rudadr} or 
positions of \cite{Sergey} condensed counterions play an essential 
role in bringing about a sudden aggregation of the DNA molecule(s).  
These interactions are quite intricate since they can be described by 
a pairwise additive potential only at very low concentrations of DNA 
\cite{rudadr},  \cite{andrea}.  For dense DNA mesophases, many-body effects 
appear to be essential \cite{andrea2} and any theory of the DNA 
condensed state ought to to take them into account.  In comparison, 
the description of the solution phase, where pairwise additive 
interactions make sense, 
promises to be much simpler and more easily within reach.

Our goal here is to assess some of the fundamental consequences that 
attractive electrostatic correlation forces have on the thermodynamic 
properties of stiff, charged polymers.  In certain respects this 
problem has already received some attention in the framework of liquid 
crystalline ordering of polyelectrolye solutions \cite{grokho}, 
\cite{nyrkova}.  Here we find it useful to focus entirely on single-chain 
properties.  A related approach was introduced recently by Golestanian 
{\sl et ~al} \cite{golest}.

We start with the mean field theory of a stiff polyelectrolyte chain, 
where thermal fluctuations are ignored; we find that counterion 
correlation attractions should induce an Euler-like buckling 
instability of the chain \cite{timosh}.  Formally, we show that at 
zero temperature, the buckling instability is associated with 
nontrivial solutions of a Shr\" odinger-like equation which is solved 
for different models of interaction potentials.  We determine how 
$V_{cr}$, the critical strength of the potential at which buckling 
sets in, depends on the length of the polyelectrolyte.  Below the 
instability, the chain is straight whereas above the instability 
point, the chain buckles from intrachain attractions.


In steps, we then treat the effect of the fluctuations on this 
mean-field picture, first, within a harmonic theory that Odijk 
\cite{odi}, in a related context, has dubbed the ``semiclassical'' 
picture.  This picture is strictly valid only when fluctuation effects 
are quantitatively small.  At this level, there are indications that 
the buckling instability is preserved but that thermal fluctuations 
tend to renormalize the value of the persistence length of the chain, 
so as to make it {\sl smaller} than the bare value.  The harmonic 
approach breaks down close to the buckling instability where 
conformational fluctuations become large.  The predictions can 
therefore only be considered as plausible trends.

The inclusion of thermal fluctuations of arbitrary strength, however,
does not conform to the  mean-field or
 the ``semiclassical'' picture.
We propose a new approach
that can describe properly the thermodynamic properties of the chain 
for any set of parameters.  This approach involves a systematic 
$1/d$-expansion \cite{David} of the partition function, where $d$ 
denotes the dimension of embedding space.  We prove that, in general, 
thermal fluctuations completely destroy the (mean-field or 
``semiclassical'') buckling instability and convert it into a polymer 
collapse \cite{grokho}.  By varying the bare persistence length, and 
the strength and range of the attractive potential,

particular, we determine a complete phase diagram.  We find that true 
buckling occurs only at zero temperature (or infinite persistence 
length) and that thermal fluctuations at any finite temperature turn 
buckling into collapse.  We do not address the nature of the collapsed 
state (as was done in e.g.  \cite{nyrkova}); we would have to include 
consistently into the the theory, many-body, non-pairwise additive 
effects in the electrostatic correlation attractions.  This, however, 
is still too difficult.


As with many concepts in polyelectrolyte theory, the idea that an
Euler-like elastic buckling instability could lie at the core of the
polymer collapse problem has been put forth by Manning in relation to
DNA condensation \cite{manni}. Manning's calculation, however, does
not take into account explicitly the attractive part of the
interaction potential along the polyelectrolyte molecule. The same is
true for the recent analysis set forth by Odijk \cite{odi}. Here we
shift the perspective  by explicitly considering the
contribution of attractive forces between polyelectrolyte segments to
the onset of a buckling instability. In addition, we show how thermal
fluctations may transform the buckling behavior into a polymer
collapse.

The exact form of the counterion fluctuation potential in the regime 
where pairwise additivity makes sense, is not of serious concern here.  
Our theory is valid for any form.  We indicate that generally by doing 
some of the calculations for different model forms of the pair 
potential.  However, general arguments based on the analysis of the 
assymptotic form of the counterion fluctuation forces indicate quite 
strongly \cite{rudadr}, \cite{belloni} that it is not unreasonable to expect 
that the effective pair interaction potential $V(u)$ is of the generic 
form $V(r) = - |v_0|{\left(\frac{1}{r}\right)}^2 e^{-\kappa r}$, where 
$r$ is the monomer-monomer separation, $a$ is roughly the size of an 
effective monomer.  $v_0$ sets the energy scale for the interaction, 
and $\kappa$ an inverse screening length.  We often invoke this form 
but it is not essential for general conclusions.

The organization of this paper is as follows. In Section 2 we present
the mesoscopic free energy of the chain and derive the elastic
equilibrium (mean-field) equation for its shape.
the
In Section 3 we investigate numerically the nature of
this instability for different model interaction potentials. Next we
turn to the effect of the fluctuations. The harmonic
(``semiclassical'') theory is considered in Section 4.
persistence
The analysis based on a  systematic $1/d$-expansion
of the partition function, is presented in Section 5.
Finally, in Section 6 we discuss our results and address the
question of possible experimental ramifications of our results.

\section{Mean-field Theory}
\begin{figure}[htb]
\epsfxsize= 8 cm
\centerline{\epsfbox{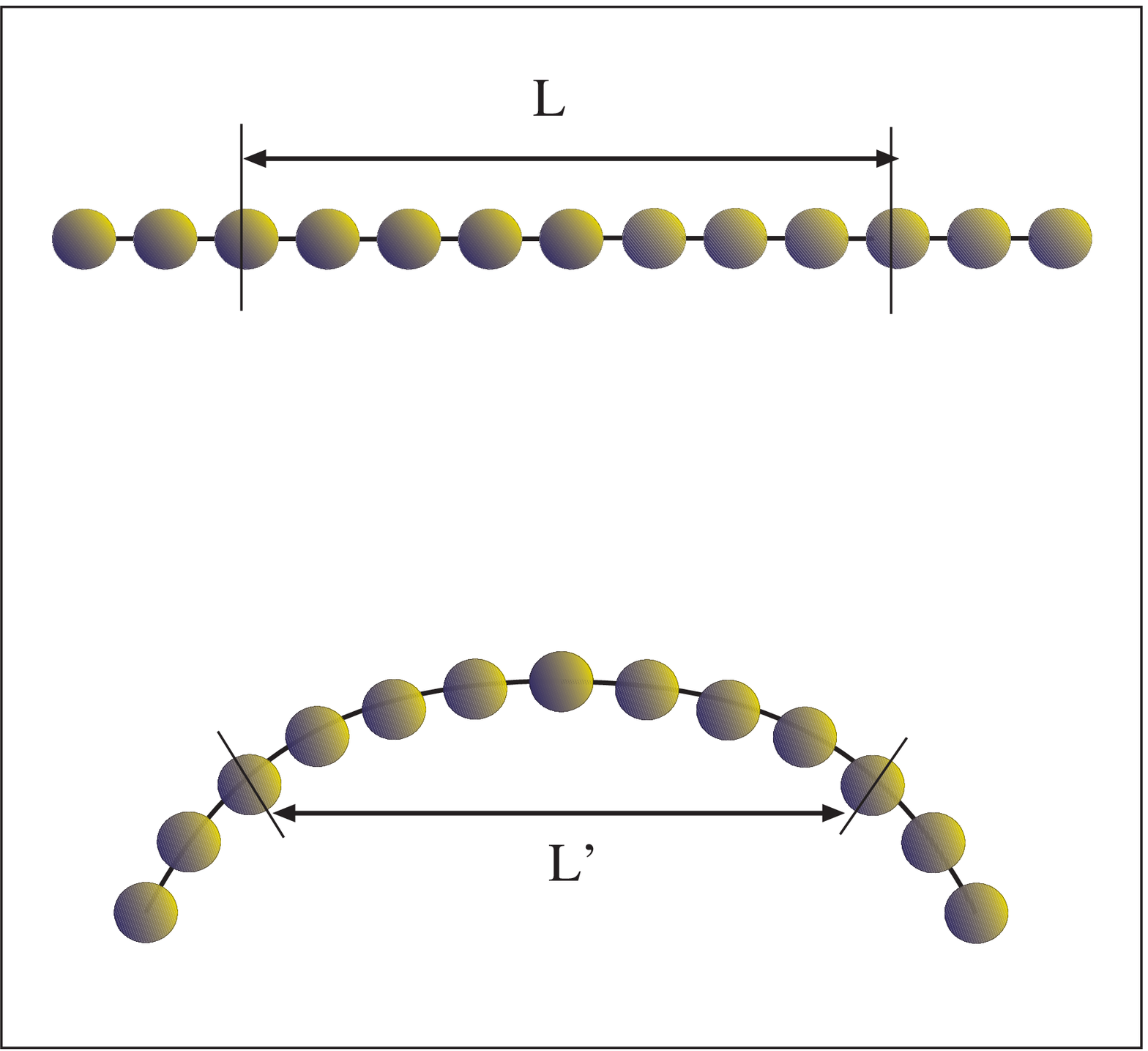}}
\caption{Schematic representation of buckling. In a buckled chain the 
effective separation between segments becomes smaller which lowers 
the free energy of the chain if the interactions between segments are 
attractive. This decrease in the free energy works against  the 
increased bending energy thus leading to an instability.}
\label{fig-1}
\end{figure}

The idea of an interaction-induced buckling instability is really
quite intuitive, see Fig. \ref{fig-1}. At the point where the elastic
energy of bending can no longer compensate for the change of the
interaction energy due to diminished separations between the
interacting
segments of the polymer chain, the polymer will buckle. Clearly,
this buckling depends
on the persistence length of the polymer as well as on the
``strength'' of the attractive interactions between its segments.

The starting point for a mean-field description of this buckling
instability is the identification of a  mesoscopic free energy of
the self-interacting persistent polymer. We suggest here that
this free energy may be written as a sum of configurational elastic
energy
 and pair-interaction energy for all the monomers:
\begin{eqnarray}
{\cal F} &=& \frac{K_c}{2} \int_0^{L}
\frac{ds}{R^2} + {\textstyle{1\over 2}}
\int_0^
{L}\!\!\int_0^{L} ds ds'~V\left(\vert{\bf
r}(s)-{\bf r}(s')\vert\right)  \nonumber\\
&=& \frac{K_{c}}{2} \int_0^{L} ds~
{\left( \frac{\partial^{2}Ä{\bf{r}}}{\partial s^{2}Ä} \right)}^{2}Ä +
\frac{1}{2}\int_0^{L}\int_0^{L} ds ds'~ V(|{\bf r}(s)-{\bf r}(s')|)~,
\label{hamiltonian}
\end{eqnarray}
where ${\bf{r}}(s)$ denotes a general parametrization of the the
polymer
chain,
$s$ the contour length, and  $|{\bf{r}}(s)-{\bf{r}}(s')|$ is the
distance
in embedding space betweeen two monomers at $s$ and $s'$.
$K_c$ is the elastic modulus, which is related to the bare
persistence
length, ${\ell}_p,$ via $\frac{K_c}{k_BT} = {\ell}_p$.
Finally,  $V\left(\vert{\bf r}(s)-{\bf r}(s')\vert\right)$ denotes
the
monomer-monomer interaction potential assumed to be purely attractive
(unless indicated otherwise). In what follows we shall remain within
the
framework of the Euler elastic rod theory, ignoring possible
non-planar configurations pertinent to the Kirchhoffian description
\cite{refkirk}.

Since we limit ourselves to considering inextensible chains,
we ought, in principle, to take into account the
constraint, $\partial_{s}{\bf{r}}(s) \cdot \partial_{s}{\bf{r}}(s)
=1$,
for any
point ${\bf r}(s)$ along the chain. For weak undulations, or, in the
limit of validity of the mean-field approximation, this constraint can
be safely ignored \cite{odi}. In general, however, one has to deal
with it appropriately, see Sec.~5.

In the mean-field approximation we determine a  typical configuration
of the polymer by functionally minimizing the free energy w.r.t.
${\bf r}(s)$. The result of such a variation is:
\begin{eqnarray}
K_c\frac{\partial^4{\bf r}(s)}{\partial s^4}+ \int ds'~{\bf{F}}
(|{\bf r}(s)-{\bf r}(s')|)=0~, \label{MF}
\end{eqnarray}
where ${\bf{F}}(|{\bf r}(s)-{\bf r}(s')|)=-\partial V/ \partial
{\bf{r}}(s)$
is the local force density.  This equation determines the typical 
polymer configuration subject to appropriate boundary conditions.  It 
is easy to show that if we consider a polymer with non-interacting 
monomers subject to the boundary conditions that the ends of the 
polymer do not bend, the solution to the mean-field eqution, 
Eq.~(\ref{MF}), corresponds to a straight rod-like configuration.  If 
we imagine that the attractive potential between monomers is weak, 
then deformations away from the rod-like configuration must be small.  
Under such circumstances, it is useful to change the parametrization 
of the polymer and to describe the polymer, and solve the mean-field 
equation, in a coordinate system which is well suited to describe 
small deformations away from the straight configuration.

Specifically, for small deviations from a straight configuration,
the polymer chain can be parameterized as ${\bf r}(s) =
\left( z,{\brho}(z)\right)$,
where $z$ is the direction of the axis of the molecule and
$|{\bf{\rho}}|$ is the radial distance from that axis. In this
parameterization one has $ds = dz \sqrt{1 +
\left(\frac{d\brho(z)}{dz}\right)^2}$ and $\frac{1}{R} =
\frac{\frac{d^2\brho(z)}{dz^2}}{\left( 1 +
\left(\frac{d\brho(z)}{dz}\right)^2\right)^{3/2}}$. With
${\brho}^{\prime}(z) = \frac{d\brho(z)}{dz}$, the free energy can
be written as follows:
\begin{eqnarray}
\kern-80pt{\cal F} &=& \int_0^{L} dz~{\cal L}\left[
{\brho}^{\prime\prime}(z), {\brho}^{\prime}(z), \brho(z), z\right]
\nonumber\\
&=& {\textstyle{1\over 2}} K_c \int_0^{L} dz
{(\brho^{\prime\prime}(z))}^2\left( 1 + {(\brho^{\prime}(z))}^2
\right)^{-5/2}  \nonumber\\
&&~~~~~~~~~+ {\textstyle{1\over 2}} \int_0^{L}\int_0^{L}
dz dz' \sqrt{1 + {(\brho^{\prime}(z))}^2}\sqrt{1 +
{(\brho^{\prime}(z'))}^2}
V\left(\vert{\bf r}(z)-{\bf r}(z')\vert\right)~.
\label{prva}
\end{eqnarray}
The Euler-Lagrange equation for this free energy reads:
\begin{equation}
\frac{\partial{\cal L}}{\partial\brho(z)} -
\frac{d}{dz}\frac{\partial {\cal{ L}}}{\partial\brho^{\prime}(z)} +
\frac{d^2}{dz^2}\frac{\partial {\cal{
L}}}{\partial\brho^{\prime\prime}(z)}
= 0~.
\label{foo}
\end{equation}
Since here we will be interested only in the limit of a
straight rod solution we can linearize Eq.~(\ref{foo}),
deformations,
where
\begin{equation}
K_c \frac{d^4\brho(z)}{dz^4} - \int_0^{L}dz' \frac{d}{dz}\left(
\frac{d\brho(z)}{dz} V(\vert{\bf r}(z)-{\bf r}(z')\vert)\right)
= - \int_0^{L}dz' \frac{\partial V(\vert{\bf r}(z) - {\bf
r}(z')\vert)}{\partial\brho(z)} \label{ELL}~.
\label{first}
\end{equation}
The first term on the l.h.s. of  Eq.~(\ref{ELL}) stems from the
curvature energy. The second is the longitudinal stress acting
along the deformed rod. The term on the r.h.s. corresponds to the
transverse bending force \cite{LL}. Both of the last two terms depend
on
the characteristics and the details of the interaction potential.

By considering
the first variation of the free energy at the boundaries, we derive
the boundary conditions for Eq.~(\ref{ELL}). Assuming that both ends
are free, {\sl i.e.}, that the
variations of $\delta\brho(z = 0,L)$ and $\delta\brho^{\prime}(z =
0,L)$
are arbitrary, we find:
\begin{eqnarray}
\brho^{\prime\prime}(z = 0,L) &=& 0 \nonumber\\
\left(\brho^{\prime\prime\prime}(z = 0,L) -
\brho^{\prime}(z = 0,L)\int_0^{L}dz'V({\bf r}(z = 0,L) -
{\bf r}(z'))\right) &=& 0 ~.
\end{eqnarray}

In the case of short range interactions, the r.h.s. term of the
Euler-Lagrange equation, Eq.~(\ref{ELL}), can be simplified further.
Significant contributions to the integral on the {\sl r.h.s.} of
Eq.~(\ref{first})
come only from the points along the polymer for which $z$ and $z'$ are
not very much apart. Therefore one can develop $V({\bf r}(z) - {\bf
r}(z'))$ in powers  of $z-z'$, truncating the Taylor expansion at the
first order term:
\begin{eqnarray}
\brho(z') - \brho(z) \simeq  \frac{d\brho(z)}{dz}(z - z') + \dots
\nonumber\\
{\bf r}(z) - {\bf r}(z') = \left( z-z', \brho(z) -
\brho(z')\right) \simeq (z-z')\left(1,\frac{d\brho(z)}{dz}\right).
\end{eqnarray}
Thus we end up with the following approximation:
\begin{equation}
\frac{\partial V(\vert{\bf r}(z) - {\bf
r}(z'))\vert}{\partial\brho(z)} \simeq \frac{\partial V(\vert
u\vert)}{\partial u} \frac{\brho^{\prime}(z) (z-z')}{\vert
z- z'\vert\left( 1 +
\textstyle{1\over 2}{(\brho^{\prime}(z))}^2 + \dots \right)} \simeq
\frac{\partial V(\vert z - z'\vert)}{\partial z} \brho^{\prime}(z) +
\dots
\end{equation}
To the lowest order in $\brho^{\prime}(z)$, the Euler-Lagrange
equation reads:
\begin{eqnarray}
K_c \brho^{(iv)}(z) - \int_0^{L}dz'
V(\vert z - z'\vert)\brho^{\prime\prime}(z) \simeq 0~\label{schroe}.
\end{eqnarray}
Eq.~(\ref{schroe}) is the fundamental mean-field equation describing
the shape
of the elastic rod in the limit of small deformations and short range
self-interactions.
It is closely related to the Schr\" odinger equation if one
introduces the
variable ${\bf u}(z) \equiv \brho^{\prime\prime}(z)$. One finds:
\begin{equation}
{\bf u}^{\prime\prime}(z) - \Omega^2(z){\bf u}(z) = 0 ~~~~~{\rm
with}~~~~~\Omega^2(z) = K_c^{-1} \int_0^{L}dz' V(\vert z - z'\vert)~,
\label{schro}
\end{equation}
and the boundary condition assumes the form:
\begin{equation}
{\bf u}(z = 0,L) = 0~. \label{b.c.}
\end{equation}
By definition, $\Omega(z)$ is a symmetric function with
respect to the midpoint of the rod.  By integrating ${\bf u}(z) =
\brho^{\prime\prime}(z)$, one also gets the part that corresponds to
$\brho^{\prime\prime}(z) = 0$, {\sl i.e.},  $\brho(z) = {\bf A} + {\bf
B}z$, corresponding to a straight undeformed rod. The field ${\bf
u}(z)$ is therefore a good measure of how much the polymer is deformed
away from the straight
rod configuration. In the Eulerian description all ${\bf u}(z)$'s are
coplanar.

The mean-field equation, Eq.~(\ref{schro}), is not easily solvable
except
for a limited variety of potentials, $V(u)$. Nevertheless, a formal
solution can be obtained by introducing the following ansatz for
$u(z)$, assumed now to lie within a single plane (Eulerian
description):
\begin{equation}
u(z) = {\cal C}\eta(z) \sin\left( \Phi(z,0)\right)~.
\end{equation}
${\cal C}$ is a normalization constant introduced in order to make
$\eta(z)$ dimensionless.
The functions $\eta(z)$ and $\Phi(z,0)$ are easily shown to satisfy
the
following set of equations:
\begin{eqnarray}
\eta^{\prime\prime}(z) + \Omega^2(z)\eta(z) -
\eta(z){(\Phi^{\prime})}^2(z,z')
&=& 0 \nonumber\\
{\left( \eta^2(z){\Phi}^{\prime}(z,z') \right)}^{\prime} &=& 0~.
\label{glupus}
\end{eqnarray}
The instability point is reached when the following identity
is fulfilled:
\begin{equation}
\frac{1}{L}\int_{0}^{L} \frac{dt}{\eta^2(t)} = \pi~,
\label{butec}
\end{equation}
which follows from the fact that the non-trivial solution should
satisfy the boundary condition Eq.~( \ref{b.c.}) which is translated into
$\sin{\Phi(L,0)} = 0$. Eq.~(\ref{butec}) selects the lowest mode
compatible with this condition.

Because the function $\eta(z)$ still has to be evaluated,
this is nothing but a formal statement of the properties of
the solution of the mean-field equation. The general properties
can be obtained from a WKB {\sl ansatz} \cite{FeynmanHibbs}.
At this level, the comparability
$\eta^{-2}(z) \sim
\Omega(z)$, leads to a stability limit described by:
\begin{equation}
\int_0^{L} dz \sqrt{\vert \int_0^{L}dz'\beta V(\vert z -
z'\vert)\vert} = \sqrt{\beta K_c} \pi~.
\label{wkbres}
\end{equation}
The necessary condition for the existence of the instability is:
\begin{equation}
\int_0^{L}dz' V(\vert z - z'\vert) < 0~.
\end{equation}
The properties of this instability (bifurcation) point are the same
as in the case of the simpler Euler instability  where the elastic rod
is simply compressed at both ends by a transverse force. This
latter case has previously been considered by Manning \cite{manni}.

\section{Mean-field Solution for Different Model Potentials}

In order to get a feel for the solutions of the mean-field equation
for the shape of the self-attracting polymer chain,
Eq.~(\ref{schro}), we
solve it for three different model interaction potentials,
V(r): A finite
potential well, an exponential potential, and a counterion correlation
potential. Approximate analytical solutions can be obtained only for
the first two potentials.

In the case of a finite potential well,
\begin{equation}
V(r) = \left\{ {\begin{array}{ll}
                 - V_0 & \mbox{if $r < r_0$} \\
                   0   & \mbox{otherwise~,}
                 \end{array}} \right.
\label{potential}
\end{equation}
which leads to,
\begin{equation}
\Omega^2(z) = K_c^{-1} \int_0^{L}dz' V(\vert z - z'\vert) =
          \left\{ {\begin{array}{lll}
                 - \frac{2V_0r_0}{K_c}        & \mbox{$r_0 < z < L -
r_0$}
\\
                 - \frac{V_0(z + r_0)}{K_c}   & \mbox{$z < r_0$} \\
                 - \frac{V_0(L -z + r_0)}{K_c} & \mbox{$L - r_0 < z <
L$~.}
                 \end{array}}
          \right.
\end{equation}
The solutions of the mean-field equation, Eq.~(\ref{schro}), with this
potential are the angular functions and the Airy functions
\cite{stegun}.
Taking into account the boundary condition and the  continuity
 of the solution
and its derivatives at the discontinuities of the potential,
Eq.~(\ref{potential}), we obtain the following approximate form for
the
critical magnitude of $V_{cr}$:
\begin{equation}
\frac{V_{cr}r_0L^2}{K_c} \sim \frac{\pi^2}{2}\left( 1 + \frac{\pi^2
r_0^3}{6~L^3} + \dots \right)~.
\end{equation}
For large enough $L$, the scaling behavior of $V_{cr}$ has already
been
determined by Manning \cite{manni}. Obviously, the scaling form
derived in
Ref.~\cite{manni} is only valid when the condition $\frac{r_0}{L} \ll
1$
is satisfied.

The next explicitly solvable model is the exponential potential,
which can be viewed as a generic form of a short range potential.
Here
\begin{equation}
V(r) = - V_0 e^{-r/r_0},
\label{exponential}
\end{equation}
where $V_0$ is a constant. For this potential
\begin{equation}
\Omega^2(z) = K_c^{-1} \int_0^{L}dz' V(\vert z - z'\vert) =
\frac{2V_0}{K_c} - \frac{2V_0}{K_c} e^{-\frac{L}{2 r_0}} {\rm
cosh}(z/r_0)~,
\end{equation}
where we have  displaced the origin of the $z$ axis to the midpoint
of the
polymer chain, {\sl i.e.}, $z \rightarrow z + \frac{L}{2}$.
Introducing the variable $2t  = z / r_0$, we obtain the mean-field
equation, Eq.~(\ref{schro}), on the form,
\begin{equation}
\frac{d^2y(t)}{dt^2} - \left( a - 2q{\rm cosh}(2t) \right)y(t) = 0~,
\label{sch}
\end{equation}
which is equivalent to the modified Mathieu equation with standard
parameters
 $a = - \frac{8V_0}{K_c{r_0^3}}$ and $q = - \frac{4V_0}{K_c{r_0^3}}
\exp{-\frac{L/r_0}{2}}$ \cite{stegun}. There is only one solution
 of this equation which satisfies the requirements of being
both symmetric with respect to the origin of $z$-axis, and finite in
the limit $q \rightarrow 0$:
\begin{equation}
y(t) \sim \sum_{n=-\infty}^{n=+\infty} \stackrel{(s)}{c_{2n}} K_{2n +
s}\left(2\sqrt{q} {\rm cosh}(t)\right)~,
\end{equation}
where $s$ is the solution of the Hill equation \cite{Hill},
\begin{equation}
\sin^2{\frac{\pi}{2}s} = - \Delta(0) \sinh^2{\frac{\pi}{2} \sqrt{a}}~,
\end{equation}
with $K_n(x)$ denoting  the modified Bessel function, and $\Delta(0)$
 the Hill determinant for $s = 0$ \cite{Hill}. This equation can be
solved
explicitly only in a limit that  would correspond to $L/r_0 \gg
1$. In this limit $\Delta (0) \simeq 1$ wherefrom:
\begin{equation}
\sin^2{\frac{\pi}{2}s} \simeq - \sinh^2{\frac{\pi}{2}
\sqrt{a}},~~\longrightarrow ~~ s \simeq \imath \sqrt{a}~.
\end{equation}
To the lowest order, {\sl i.e.}, for $n=0$, the solution of the Schr\"
odinger equation reads:
\begin{equation}
y(t) \sim K_{\imath \sqrt{a}}\left(2\sqrt{q} \cosh{t}\right)~.
\end{equation}
The boundary condition at $t = \pm \frac{L/r_0}{2}$ thus comes out
as
\begin{equation}
K_{\imath \sqrt{a}}\left(\sqrt{\frac{a}{2}}\left( 1 +
e^{-\frac{L/r_0}{2}}\right)\right) \simeq 0~,
\end{equation}
and has to be determined for $a = \vert \frac{V_0}{2 r_0{\cal
K}}\vert$. Only the asymptotic form of the solution can be obtained
explicitly, as follows:
\begin{equation}
\frac{V_{cr}r_0L^2}{K_c} \sim \frac{\pi^2}{2}\left( 1 +
4\pi^2 \frac{ r_0^3}{L^3} + \dots \right)~,
\label{result}
\end{equation}
not unlike the result for the finite potential well. The details of
the interaction potential thus, at least in the asymptotic limit, do
not appear to matter much. Clearly, the asymptotic form derived by
Manning
\cite{manni} again provides a reasonable description of the point of
 instability,
the lowest order deviation from it varying as
$\left(\frac{r_0}{L}\right)^3$.

\begin{figure}[htb]
\epsfxsize= 8 cm
\centerline{\epsfbox{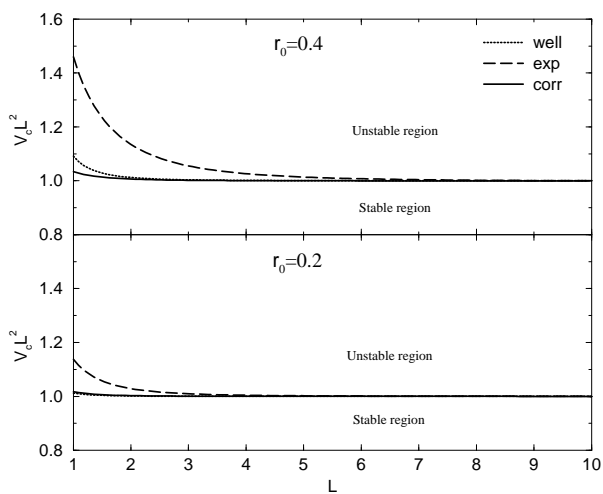}}
\caption{Critical strength of the interaction potential, in terms of
$V_{cr}~L^2$, as a function of the length of the rod, $L$, at
different values of the screening parameter $r_0$ for the three models of attractive
potentials: Finite potential well, Eq.~(\ref{potential}), exponential
potential, Eq.~(\ref{exponential}), and the general correlation
potential, Eq.~(\ref{correlation}). The dependencies have been rescaled in such a
way that at large $L$ they coincide.}
\label{fig0}
\end{figure}

The last form of the fluctuation potential that we consider is the
asymptotic form of the effective pairwise additive form of the
counterion correlation potential derived in Refs. \cite{rudadr},
\cite{belloni}:
\begin{equation}
V(r) = - V_0 a^2 \left(\frac{e^{-r/r_0}}{r}\right)^2 \equiv
        - |v_0| \frac{e^{-\kappa r}}{r}.
\label{correlation}
\end{equation}
fluctuation
There is no simple analytical result that one can derive for this
interaction potential even in the asymptotic limit. The numerical
solution is, however, revealing enough, see
Fig. \ref{fig0}. Apparently the finite range of the potential has even
less effect on the stability limit than in the previous two cases for
equal $r_0$. The stability limit is extremely well approximated by the
WKB {\sl ansatz} Eq.~(\ref{wkbres}) which suggests that the critical
strength of the attractive potential inducing buckling should be
inversely porportional to the length of the chain squared.

Comparing the three model potentials one sees that the effect of the
finite range of the potential is most pronounced for the exponential
potential and least pronounced for the correlation form,
Eq.~(\ref{correlation}). At the mean-field level, the detailed form
of
the attractive part of the intrachain potential is thus of minor
importance
for a long enough chain.

\section{Fluctuations: ``Semiclassical Theory''}

The mean-field theory analyzed above completely ignores the effects of
thermal fluctuations on the conformational properties of the chain. To
include these effects at the most primitive level, we now evaluate the
partition function of the chain if the mesoscopic Hamiltonian is
expanded up to  second order in the fluctuations around a straight
rod-like configuration. By analogy with quantum mechanics,
 Odijk \cite{odi} has
dubbed this type of approach the ``semiclassical'' theory of buckling.
The fluctuations are treated very approximately in this scheme
and as soon as they become large enough, the whole approximation
breaks down. This happens, of course,  right at the
instability. Still we will argue that this generalization of the
mean-field formalism will give us trends concerning the
conformational properties of the chain close to the buckling
transition.

In order to treat the effect of the fluctuations on the stability
properties of the self-interacting polymer chain, we first of all
write
down the free energy of the chain subject to  small conformational
fluctuations. An expansion is performed on the basis of
Eq.~(\ref{prva}),
in the limit of
$\brho^{\prime}(z) \ll 1$, with the result:
\begin{eqnarray}
{\cal F} &\sim& {\textstyle{1\over 2}} K_c \int_0^{L} dz~
{(\brho^{\prime\prime}(z))}^2 + {\textstyle{1\over 2}}
\int_0^{L}\int_0^{L}
dz dz'~ V\left(\vert z - z'\vert\right)  \nonumber\\
& & + {\textstyle{1\over 4}} \int_0^{L}\int_0^{L} dz dz' \left(
{(\brho^{\prime}(z))}^2 + {(\brho^{\prime}(z'))}^2 \right)
 V\left(\vert z - z'
\vert\right) + \dots
\end{eqnarray}
This harmonic form of the free energy we now write with a new
independent variable $\brho^{\prime}(z) = {\bf{w}}(z)$ on the form,
\begin{equation}
{\cal F}({\bf{w}}(z), {\bf{w}}^{\prime}(z),z) \sim {\cal F}_0 +
{\textstyle{1\over 2}}
K_c \left( \int_0^{L} dz
~{({\bf{w}}^{\prime}(z))}^2 + \int_0^{L} dz~\Omega^2(z) {\bf{w}}^2(z)
 \right)~,
\label{action}
\end{equation}
with
\begin{equation}
{\cal F}_0 = {\textstyle{1\over 2}} \int_0^{L}\int_0^{L}
dz dz' ~V\left(\vert z - z'\vert\right)~,
\end{equation}
and
\begin{equation}
\Omega^2(z) = K^{-1}_c\int_0^{L} dz'~ V\left(\vert z -
z'\vert\right)~.
\end{equation}
The propagator for a harmonic action given by the Hamiltonian,
Eq.~(\ref{action}), can be evaluated analytically \cite{lawande} and
is determined by quantities characterizing the ``classical''
solution, as calculated via the Euler - Lagrange equation that one
can
derive
from the free energy (or ``Lagrangian''), Eq.~(\ref{action}).
In terms of the local curvature,
${\bf{w}}^{\prime}(z)$,
the propagator can be derived on the form:
\begin{eqnarray}
\kern-50pt{\cal K}({\bf{w}}^{\prime}(z),z; {\bf{w}}^{\prime}(z'), z')
 &=& \int\dots\int {\cal
D}{\bf{w}}(z)~e^{-\beta {\cal
F}({\bf{w}}^{\prime}(z),{\bf{w}}^{\prime}(z);z)}
 \nonumber\\ &=&
{\rm det}^{-1/2}\left( \frac{\partial^2\beta
F_{cl}({\bf{w}}^{\prime}(z),
{\bf{w}}^{\prime}(z');z, z')}
{\partial
{\bf{w}}^{\prime}(z) \partial {\bf{w}}^{\prime}(z')} \right)
\exp{\left( - \beta F_{cl}({\bf{w}}^{\prime}(z),
{\bf{w}}^{\prime}(z'); z, z') \right)},\nonumber\\
~
\end{eqnarray}
where $F_{cl}({\bf{w}}^{\prime}(z), {\bf{w}}^{\prime}(z'); z, z')$ is
the ``classical'' contribution to the free energy,
Eq.~(\ref{action}),
evaluated for the ${\bf{w}}(z)$ which is a
solution of
the Euler - Lagrange equation, Eq.~(\ref{schro}).  The propagator can
now
be written on closed form \cite{grothaus}:
\begin{eqnarray}
{\cal K}({\bf{w}}(z),z; {\bf{w}}(z'), z') &=&
\left( \frac{\sin{\Phi(z,z')}}{\rho(z)\rho(z')}
 \right) \times \nonumber\\ & & \kern-160pt\exp{\left( -\frac{\beta
K_c
\cos{\Phi(z,z')}}{2 \sin{\Phi(z,z')}}
L \left( {\rho^2(z)}{({{\bf{w}}^{\prime}(z))}^2}
+ {\rho^2(z')}{({{\bf{w}}^{\prime}(z'))}^2}\right) + {\beta K_c L
{\bf{w}}^{\prime}(z){\bf{w}}^{\prime}(z')}
\frac{\rho(z)\rho(z')}{\sin{\Phi(z,z')}} \right) }~, \nonumber\\ ~
\label{putz}
\end{eqnarray}
where $v_1(z) = \rho(z) \sin\Phi(z,z')$ and $v_2(z) = \rho(z)
\cos\Phi(z,z')$ are just two linearly independent solutions of
Eq.~(\ref{sch}).  The function $\rho(z)$ in Eq.~(\ref{putz}) is a
solution of the Ermakov-Pinney equation \cite{kleinert},
\begin{equation}
\rho^{\prime\prime}(z) + \Omega^2(z)\rho(z) - \frac{1}{\rho^3(z)} =
0~,
\label{pinney}
\end{equation}
while $\Phi(z,z')$ can be derived on the form
\begin{equation}
\Phi(z,z') = \frac{1}{L}\int^{z}_{z'}\frac{dt}{\rho^2(t)}~.
\end{equation}
All this follows directly from  Eq.~(\ref{glupus}). All other
derived quantities can now be obtained with the help of this local
 curvature propagator.

The density distribution function
$P[{\bf{w}}^{\prime}(z)]$ for the curvature is obtained in a straight
forward way
from the propagator. Assuming  free boundary conditions,
$\brho^{\prime\prime}(z = 0,L) = {\bf{w}}^{\prime}(z = 0,L) = 0$, one
obtains the
curvature density distribution function as follows:
\begin{equation}
P[{\bf{w}}^{\prime}(z)] \sim \exp{\left( -\frac{\beta K_c L}{2} \left(
\cot{\Phi(0,z)} +
\cot{\Phi(z,L)}\right)\rho^2(z){({\bf w}^{\prime}(z))}^2\right)}~.
\end{equation}
The normalization constant is irrelevant as we will be only interested
in the average of ${({\bf{w}}^{\prime}(z))}^2$.  Evaluating now the
local curvature fluctuations at the midpoint of the polymer chain,
$z = L/2$, we are left with
\begin{equation}
\langle {({\bf{w}}^{\prime}(L/2))}^2 \rangle = \frac{\int dy~ y^2
P[y(z=L/2)]}{\int dy P[y(z=L/2)]} = \frac{1}{\beta K_c L \rho^2(L/2)}
\tan\left({\frac{1}{L}\int_{0}^{L} \frac{dt}{2\rho^2(t)}}\right)~,
\label{usq}
\end{equation}
where $\langle \ldots \rangle$ indicates thermal averaging.
The effect of the fluctuations in the harmonic limit can now be
assessed as follows. Clearly, at the instability
$\langle {({\bf{w}}^{\prime}(L/2))}^2
\rangle$ should become large. Just how large is difficult to see
from Eq.~(\ref{usq}) since the derivation is valid only in the limit of
small fluctuations. Nevertheless, following Odijk's reasoning
\cite{odi}, we claim that the instability sets in as soon
as the relative fluctuations in the midpoint become larger than ${\cal
O}(1)$, $i.e.,$
${\langle{({\bf{w}}^{\prime}(L/2))}^2 \rangle}{\rho^2(L/2)}
= c_0 = {\cal O}(1)$. It is difficult to say more than that without
actually solving the fundamental equation, Eq.~(\ref{pinney}). Close to
the instability point, the lowest order contribution to the
 formal solution of the above equation reads:
\begin{equation}
\rho^2(L/2)\langle {({\bf{w}}^{\prime}(L/2))}^2 \rangle \sim
\frac{1}{\beta K_c L~\left(\frac{\pi}{2} - \frac{1}{L}\int_{0}^{L}
\frac{dt}{2\rho^2(t)}\right)}~.
\end{equation}
Within the WKB approximation this result has a very interesting
interpretation. Here the above instability limit can be written on the
following form, with explicit dependence on the interaction potential,
\begin{equation}
\int_0^{L} dz \sqrt{\vert \int_0^{L}dz' \beta V(\vert z -
z'\vert)\vert} \sim \pi \sqrt{\beta K_c}\left( 1 -
\frac{1}{\beta K_c
L}\frac{1}{\rho^2(L/2)\langle{({\bf{w}}^{\prime}(L/2))}^2
\rangle}\right)~.
\end{equation}
If we compare this result with the pure mean-field result,
Eq.~(\ref{wkbres}),
which excludes any effect of fluctuations, the instability point is
obviously
reached when the strength of the potential reaches the same value
as one finds in the mean-field case,
except now the value of the elastic constant is renormalized
according to:
\begin{equation}
\beta K_c \longrightarrow \beta K_c - 2\frac{1}{\rho^2(L/2)
\langle {({\bf{w}}^{\prime}(L/2))}^2 \label{attrenorm}
\rangle}~.
\end{equation}
This renormalization of the elastic constant is obviously
fluctuational and is thus linear in temperature.  Clearly, the above
reasoning is not quantitatively valid since we are stretching the
harmonic theory into a regime where it is not valid. However, one can
hope that it bears out the correct tendencies for the behavior of this
system.

It thus appears that the ``semiclassical `` theory of buckling would
lead to the same type of instability as the mean-field theory, but  with
the persistence length or, equivalently, the elastic modulus taking on a
smaller value than the bare value. In other words, thermal fluctuations
appear to renormalize the persistence length to a smaller value  than the
bare value. This is exactly the opposite of what happens in the case of
purely repulsive segment-segment interactions \cite{Odijk}.

How much of this scenario remains valid in the case where fluctuations
can not be dealt with as a small perturbation, but are essential to the
behavior of the system? The answer to this question presupposes the
complete solution of the statistical mechanical problem of the
self-interacting stiff polymer chain. This solution can be
obtained only in an approximate form such as we derive below.

\section{Fluctuations: Systematic $1/d$-expansion}
In order to treat the fluctuations on an appropriate level one has to
go beyond the simple minded approximate harmonic or
``semiclassical'' theory we described above.
In this section we will briefly
outline one  approach that goes beyond the ``semiclassical'' theory.
We wish, in particular, to introduce a
program which allows for a reasonably straight forward, approximate
calculation of the partition function and free energy for a
semi-flexible polymer whose monomers interact via a pair-potential.
The formalism we develop has already been applied to describe the
conformations and thermal properties of other intrinsically flexible
materials, including membranes. Thus, the formalism has been used to
predict the conformational behavior of fluid membranes
\cite{David&Guitter} and tethered manifolds, with
\cite{LeDoussal}, \cite{Palmeri&Guitter} and without \cite{Kadar&Bakskone}
long-range monomer interactions. In a recent study
\cite{Thirumalai&Ha} of semi-flexible polymers with non-interacting
monomers ($V=0$) an approach, similar to the one introduced here, was
used.

For the chains under considerations, we wish
to reintroduce the general parametrization, ${\bf{r}}(s)$, and
we wish explicitly to enforce the constraint of ``inextensibility'',
$ \partial_{s}{\bf{r}}(s) \cdot \partial_{s}{\bf{r}}(s) =1$. The
form of the Hamiltonian we shall prefer to use is then, cf.
Eq.~(\ref{hamiltonian}),
\begin{eqnarray} {{\cal{H}}}= \int ds~ \frac{K_{c}}{2}
{\left( \frac{\partial^{2}Ä{\bf{r}}}{\partial s^{2}Ä} \right)}^{2}Ä +
\frac{1}{2}\int\int ds ds'~ V(|{\bf r}(s)-{\bf r}(s')|)~.
\end{eqnarray}
The partition, itself, is
the path-integral over polymer  conformations weighted by
the Boltzmann weight, $\exp{( -\beta {\cal{H}})}$,
\begin{eqnarray} Z= \int {\bf{D}}[{\bf{r}}(s)]~
\prod_{s}~ \delta^{3}(
\partial_{s}{\bf{r}}(s) \cdot \partial_{s}{\bf{r}}(s) -1)~
 \exp{( -\beta
{\cal{H}} )}~~,
\end{eqnarray}
where $\beta=1/k_{B}ÄT$.  The functional
$\delta$-function guarantees that the integral involves only such
configurations that satisfy the condition of ''inextensibility''.

Two problems complicate the evaluation of the partition
function.  The first is imposed by the
functional $\delta$-function and the constraint of
``inextensibility'' which requires us to include in the
sum over polymer conformations, ${\bf{r}}$, only those
for which the tangent vectors, ${\bf{t}}=\partial_{s}{\bf{r}}$,
lie on a unit sphere. The second  problem which
complicates the  evaluation of the partition function is
the fact that, rather than being a simple quadratic (Gaussian)
form, the inter-monomer  interaction potential is,
in realistic situations, a complicated, non-local function.
A systematic way of addressing these problems takes
advantage of a ``Lagrange multiplier'' technique.
Thus, for instance,
one can enforce the constraint of ``inextensibility'',
if one introduces an auxillary field or ``Lagrange multiplier'',
 $\lambda(s)$,
and adds to the Hamiltonian the term,
\begin{eqnarray}
\delta{\cal{H}}_{1}Ä = \frac{1}{2}\int d{s}~\lambda(s)(~{(
\partial_{s}{\bf{r}}(s) )}^{2}Ä-1 ~)~.
\end{eqnarray}

Similarly, in order to avoid the complicating non-local
term in the pair-potential, one can introduce the independent field
$B=B(s,s')$, and make the replacement
$V({({\bf{r}}(s)-{\bf{r}}(s'))}^{2}) \rightarrow V(B)$.
In order to be able to make this replacement
in a systematic way, one must somehow
enforce the constraint ${({\bf{r}}(s)-{\bf{r}}(s'))}^{2}Ä=B $.
One can do that via yet another auxillary field
(``Lagrange multiplier'')  \cite{LeDoussal}, \cite{Palmeri&Guitter} and
one is thus led to introduce
another term in the Hamiltonian,
\begin{eqnarray}
\delta{\cal{H}}_{2} = \frac{1}{2} \int
d{s}ds'~ g(s,s')(~{({\bf{r}}(s)-{\bf{r}}(s'))}^{2}Ä-B(s,s')~)~~.
\end{eqnarray}
Given these modifications, the evaluation of the
partition function now involves a much easier,
unconstrained summation over polymer conformations ${\bf{r}}$.
The price one has to pay for this simplification is that, in addition to
summing over
${\bf{r}}$, one must now sum over $\lambda$, $B$,
and $g$ as well:
\begin{eqnarray}
Z= \int {\bf{D}}[{\bf{r}}(s)] {\bf{D}}[\lambda(s)]
{\bf{D}}[g(s,s')]{\bf{D}}[B(s,s')]
\exp{\left( -\beta ( {\cal{H}}+
\delta{\cal{H}}_{1} +\delta{\cal{H}}_{2}Ä )  \right)}~~,\label{pf2}
\end{eqnarray}
In the expression for the partiotion function, Eq.~(\ref{pf2}),
it is understood that the summation over
$\lambda$ and $g$ are over contours that begin at $-i\infty$ and
end at $+i\infty$.



It is easy to see that the introduction of
``Lagrange multipliers'' provides us with an expression for the
partition function
which is quadratic in $\bf{r}$ and therefore {\it exactly} solvable as for
the integration over polymer conformations.  If one fixes
$ \Psi=\{ \lambda, B, g \}$ and expands about a particular reference
configuration, ${\bf{r}}={\bf{r}}_{0}Ä$, which has the property of
minimizing ${\cal{H}}_2={\cal{H}}+
\delta{\cal{H}}_{1} +\delta{\cal{H}}_{2}Ä $, i.e.,
 $\delta {\cal{H}}_2[{\bf{r}}] / \delta {\bf{r}} |_0 =0$,
then one finds, after integration,
an effective Hamiltonian,
\begin{eqnarray}
 {\cal{H}}_e[{\bf{r}}_{0},\Psi]=
      {\cal{H}}_2[{\bf{r}}_{0}, \Psi]
       +{ k_{B}T}\frac{d}{2}Ä
          {\rm Tr} \ln
           \left(~
              \delta(s-s')(~K_{c}Ä \partial^4_s -\partial_{s}Ä\lambda
                    \partial_{s}Ä~) + 2g_{c}Ä(s,s')
            ~\right)~,
\end{eqnarray}
where,
\begin{eqnarray}
g_{c}Ä(s,s')= g(s,s')-\frac{1}{2}\delta(s-s')\int
ds''(g(s,s'')+g(s'',s'))~,
\end{eqnarray}
and $d$ is the number of components of the vector $\bf{r}$ or,
equivalently, the dimension of embedding space.
If one ignores end effects (by considering a closed polymer, or by
enforcing periodic boundary conditions, say),
one can assume that $\lambda$ is a constant, and that $B(s,s')=B(s-s')$,
 $g(s,s')=g(s-s')$.
It is then possible to perform the diagonalization in terms of
Fourier modes, so that
\begin{eqnarray}
 {\rm Tr} \ln (~\ldots~)
        &\rightarrow & \int ds \int \frac{dq}{2\pi} \ln (~\ldots~)
        \nonumber \\
 K_c \partial^4_s -\lambda \partial_{s}^2
        &\rightarrow&  K_{c}Ä q^{4}Ä +\lambda q^{2}Ä \label{transf}
        \\
 g_{c}Ä(s,s')
        &\rightarrow &  g(q)-g(q=0)Ä~~,
        \nonumber
 \end{eqnarray}
where $g(q)Ä$ is the Fourier transform of $g(s-s')$. The calculation
of ${\cal{H}}_e[{\bf{r}}_{0},\Psi]$ is then, obviously, straight forward.

What remains in the calculation of the partition function and the
free energy, is
the more difficult integrations over $\lambda$, $B$, and $g$. These
integrations can not be performed exactly in the general case.
If, however, $d\rightarrow \infty$, the integrals are completely
dominated by the contributions from the saddle point, obtained by
minimizing
w.r.t. $\lambda$, $B$, and $g$. In this limit, the {\it exact}
expression for the free energy of the reference configuration,
${\bf{r}}_{0}Ä$, is therefore
\begin{eqnarray}
 F [{\bf{r}}_{0}]= F_{0}Ä+
      \left(
        {\cal{H}}_2[{\bf{r}}_{0},\Psi]
       +k_{B}T\frac{d}{2}Ä
         {\rm Tr }\ln \left(~  K_{c}Ä\partial^4_s -\lambda
               \partial^{2}_{s} + 2g_{c}Ä(s,s') ~  \right)
       \right)_{{SP}}Ä ~~,
\end{eqnarray}
where $F_{0}Ä$ is an unimportant constant. $SP$ implies
that the expression is evaluated at the saddle
point and the ${\rm Tr} \ln$-part can be evaluated with the help of
Eq.~(\ref{transf}). For finite $d$, corrections
to the saddle point estimate will be of order $o(d)$
and may be calculated via a {\it systematic} $1/d$-expansion
\cite{David}.
We shall not do so, being content with the  calculation
by the saddle point method.  It is possible to show that
this approximation is equivalent to relaxing the local
constraints,
$\partial_{s}{\bf{r}}(s)\cdot \partial_{s}{\bf{r}}(s)=1$
and
${({\bf{r}}(s)-{\bf{r}}(s'))}^{2}=B(s,s')$,
and replacing them by the global constraints
$ \langle \partial_{s}{\bf{r}}(s)\cdot \partial_{s}{\bf{r}}(s)
\rangle =1$,
and
$\langle {({\bf{r}}(s)-{\bf{r}}(s'))}^{2}Ä\rangle =B(s,s')$
\cite{Thirumalai&Ha}, \cite{Polyakov}.

We can now carry out a more quantitative discussion of
the properties of semi-flexible polymers with pair-wise
monomer interactions, which are here taken to be attractive.
 Within the
formalism described above, such a discussion
can, unfortunately, only be performed for simple choices of
reference configurations, ${\bf{r}}_{0}Ä$. Here we shall
confine our-selves to the choice ${\bf{r}}_{0}Ä=\zeta
s {\bf{e}}$, where $\bf{e}$ is a one-dimensional unit vector,
and  $\zeta$ is a ``stretching factor'' \cite{David&Guitter1},
whose nature will be described below.

It turns out that very  useful information is contained
in the saddle point equations and we shall analyze
these in some detail.
By functionally minimizing w.r.t. $\lambda$, $B(s-s')$, and
$g(s-s')$, one finds after some manipulations:
\begin{eqnarray}
1&=&  \frac{ \partial{\bf{r}}_{0} }{ \partial s }\cdot
      \frac{ \partial{\bf{r}}_{0} }{ \partial s }
       +    dk_{B}TÄ\int\frac{dq}{2\pi}
           \frac{ q^{2}Ä}{K_{c}Ä q^{4}Ä
         +\lambda q^{2}Ä + 2g_{c}(q)} \label{sp1} \\
g_{c}(q)Ä &=& \int ds ~(1-\cos(qs)) V'(B(s)) \label{sp2}\\
B(s-s') &=& {( {\bf{r}}_{0}Ä(s)-{\bf{r}}_{0}Ä(s') )}^2 +
         2dk_{B}TÄ\int\frac{dq}{2\pi}\frac{1-\cos(q(s-s'))}{ K_{c}Ä q^{4}Ä
         +\lambda q^{2}Ä + 2g_{c}(q)} \label{sp3}~,
\end{eqnarray}
where $V'(z)=\partial_{z}ÄV$. These equations are special
cases of more general equations obtained
by Le Doussal \cite{LeDoussal} and by Palmeri and Guitter
\cite{Palmeri&Guitter} in their analysis of elastic manifolds with
long-range monomer interactions. Of these equations, the
first, Eq.~(\ref{sp1}), guarantees
that the constraint,
$\partial_{s}{\bf{r}}(s)\cdot \partial_{s}{\bf{r}}(s)=1$,
is satisfied globally, and the third equation, Eq.~(\ref{sp3}),
takes care of the
constraint, ${( {\bf{r}}(s)-{\bf{r}}(s'))}^{2}Ä=B(s,s')$. Finally,
the second equation, Eq.~(\ref{sp2}),
determines an effective ``self-energy'' of the
polymer. This ``self-energy'' may be  expanded in
(even) powers of $q$, and the expansion coefficients determine
contributions to the renormalized elastic constants, as may
be seen from the expression for the  ``propagator''
$K^{-1}(q)=1/( K_{c}Ä q^{4}Ä+\lambda q^{2}Ä + 2g_{c}(q) )$.
Roughly speaking, the expansion coefficients tell
how the {\it non-local} interactions modify the parameters
involved in a {\it local} description of the polymer. In particular,
V'(B(s)), will, in part, determine a contribution to the
total, renormalized bending rigidity, as may be seen by analyzing
Eq.~(\ref{sp2}) (see further below).

If, in addition to minimizing w.r.t $\Psi$,
one minimizes w.r.t $\zeta$,
so as to determine the best choice of configuration in the class
of configurations defined by the equation ${\bf{r}}_{0}Ä=\zeta
s {\bf{e}}$, one finds, in agreement with Refs. \cite{LeDoussal},
\cite{Palmeri&Guitter},
\begin{eqnarray}
\lambda^{(R)}Ä=\partial_{q}^{2}K(q)|_{q=0}Ä=0~~~{\rm or}~~~\zeta=0~~,
\label{sp4}
\end{eqnarray}
where $\lambda^{(R)}Ä$ is a renormalized ``Lagrangian multiplier''.
If $\zeta\neq 0$,
the first equation in Eq.~(\ref{sp4}) expresses that
straight semi-flexible polymer
is in a stress-free configuration
(if the polymer were subjected to external
stress, the applied stress and  $\lambda^{(R)}Ä $ would
have to balance). If, on the other hand, $\zeta=0$, typical
conformations of the polymer will deviate significantly from the
straight configuration, and we may expect strongly wrinkled
configurations to dominate, or the polymer to be in a collapsed
state (see below).

It is interesting first to analyze the possibility of having a truly
straight configuration, with $\zeta\neq 0$, as the equilibrium
configuration of the polymer. Such a straight  configuration can exist
 at $T=0$, even if the interactions are attractive.
However, at non-zero temperatures we expect that only configurational
entropy, which is expected to favor random-walk behavior, can prevent
the polymer from collapsing and the polymer will,
irrespective of whether it collapses or not, have equilibrium
conformations that deviate significantly from a
rod configuration. Indeed,
one finds that for $\zeta \neq 0$, the
integral in Eq.~(\ref{sp1}) is disturbed by infra-red divergences which
can  only be removed in the special case $K_c=\infty$. This result
must be seen as indicating  that the straight-rod configuration is
unstable.
As a consequence, in the thermodynamic limit, a phase charaterized
 by a  straight average configuration
(the ordered phase) exists only in the limit $T=0$
(or $K_c=\infty$) and is otherwise
destroyed by thermal fluctuations.

Further analysis of the
conformational properties of the chain for $T>0$ must rely on a more
detailed analysis of the saddle point equations,
Eqs.~(\ref{sp1})-(\ref{sp4}).
What can we expect from such an analysis ?
As we have already indicated, at finite temperatures, we
expect that polymers whose monomers
attract have to compromise  between direct interactions
which, at short scales, favor bending and, at larger scales, favor
collapse  of the chain in order for monomers to be close, and the
different effects which counteract these processes, namely
the initial bending rigidity and conformational entropy.
If this picture is correct then we must, in agreement with
the semiclassical analysis of Sec.~4,
expect to find the effective bending rigidity, and
the persistence length, to decrease significantly, and we must expect the
theory to signal that collapse of the chain is favorable for strong enough
interactions between monomers.  The quantitative analysis confirms these
intuitive
considerations. We focus on the  change of the
bending rigidity, which may be obtained by expansion
of Eq.~(\ref{sp2}),
\begin{eqnarray}
\delta K_{c}Ä = -2 \int_{-\infty}^{\infty} ds
                 ~\frac{s^{4}Ä}{4\!} V'(B)~. \label{rigid}
\end{eqnarray}
Knowing $\delta K_c$, we  can calculate the total, renormalized
rigidity as
 $K_c^{(R)}=K_c+\delta K_c$.

It is instructive first to
consider the situation near $T=0$\cite{NOTE}.
At $T=0$ we may assume $\zeta=1$ and for the interaction
potential, $V(r)=-\frac{|v_{0}|}{r^2}Ä\exp{(-\kappaÄr)}Ä$, we find,
\begin{eqnarray}
 \delta K_{c}Ä  = 2 \frac{-|v_{0}|}{2}\int_{0}^{\infty}
                     \frac{ds}{12}
                      e^{-\kappa s}Ä(\kappa s+1)
                 = -\frac{1}{6}\frac{|v_0|}{\kappa} ~. \label{dkap1}
\end{eqnarray}
It is tempting to suggest that this significant reduction in the
rigidity could signal that the
straight rod-like configuration could become unstable even at $T=0$.
As we saw in the previous sections (see in particular
Secs.~2-3) such an instability, namely the buckling instability, does
appear.

Now, at finite temperatures, $T>0$
(and finite values of the "bare" rigidity, $K_{c}Ä$),
the $T=0$ estimate  will not describe the
conformational properties of the chain very well. We must
now take $\zeta=0$, while enforcing the constraint
of inextensibility by requiring
$\lambda^{(R)}Ä$ to assume a non-zero value. In the following
we shall assume that in analyzing $g_{c}(q)$,  it is sufficient to
retain terms up to and including the fourth order term in the Taylor
expansion of $g_{c}Ä$ in powers of $q$. This is expected to be a
valid assumption as long as $\lambda$ and $K_{c}Ä$ are not both small.
 We then find $g_{c}Ä\simeq (\delta \lambda) q^{2}Ä+
(\delta K_{c}Ä) q^{4}Ä$, where $\delta \lambda$ is the
contribution to the ``Lagrange multiplier''  from non-local
interactions, and $\lambda^{(R)}=\lambda +\delta \lambda$.
With this assumption being  made, it is easy to solve
Eqs.~(\ref{sp1}) and (\ref{sp3}), with the result:
\begin{eqnarray}
\lambda^{(R)}Ä&=&
    \left(\frac{dk_{B}T}{2}\right)^{^{2}}\frac{1}{K_{c}^{(R)}Ä}
     \label{sol1} \\
B(s) &=&  \frac{dk_{B}ÄT}{2K_{c}^{(R)}} \left( \xi^{2}Ä s +
                  \xi^{3}(e^{{-s/\xi}}-1 )\right)~, \label{sol2}
\end{eqnarray}
where $\xi=\sqrt{K_{c}^{(R)}Ä/\lambda^{(R)}Ä}$ is a cross-over
length which, in the case analyzed here, reduces to
$\xi=\ell_{p}^{(R)}/\sqrt{d/2}$.
We see that as long as there exists a self-consistent solution for
$\delta K_{c}$, the correlation function  $B(s-s') =
\langle {( {\bf{r}}(s)-{\bf{r}}(s') )}^{2}Ä \rangle$ behaves as
$\langle {( {\bf{r}}(s)-{\bf{r}}(s') )}^{2}Ä \rangle
\propto (\xi/\ell_{p}^{(R)}) |s-s'|^{2}Ä \propto |s-s'|^{2}Ä$
for small values of
$s-s'$, reflecting that the polymer is ``straight'' on short scales. We
also see that $\langle {( {\bf{r}}(s)-{\bf{r}}(s') )}^{2}Ä \rangle
\propto (k_{B}ÄT/\lambda^{(R)}Ä)|s-s'|$, for large $s-s'$, signaling
random-walk behavior.

A self-consistent solution for $\delta K_{c}Ä$, may be obtained
from Eq.~(\ref{rigid}) after inserting the solution to the saddle
point equations  Eqs.~(\ref{sol1}) and (\ref{sol2}).
One derives an integral equation whose evaluation, is complicated,
for instance, by the cross over between the rigid-rod regime
and the random-walk regime. In order to obtain the full
solution to the problem, a numerical study of the integral
equation must be carried out.
If one is satisfied with qualitative/asymptotic results, one
can analyze the integral using the method of steepest descent.

\begin{figure}[h]
\epsfxsize= 8 cm
\centerline{\epsfbox{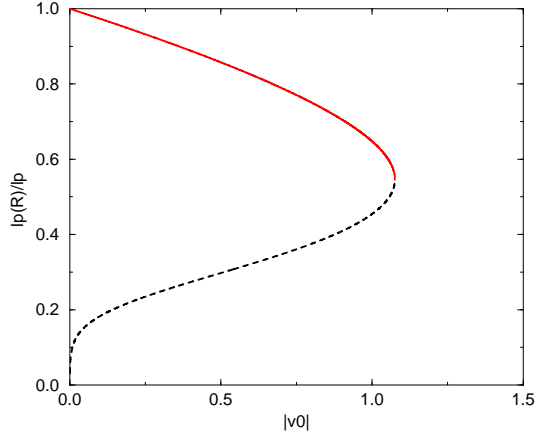}}
\caption{The solution of Eq.~(\ref{rigid}), the relation between
the strength of segmental attraction and the change in apparent
bending modulus, for $\kappa {\ell}_p=0.1$. Observe that
${\ell}_{p}^{(R)}/{\ell}_p$ depends linearly on $|v_0|$ for small
values of $|v_0|$. There exists a $|v_{0c}|$ where there is no longer a physically acceptable solution
for ${\ell}_{p}^{(R)}/{\ell}_p (|v_0|)$, implying a loss of stability 
of the coiled configuration of the polyelectrolyte chain. It is worth 
observing that $|v_{0c}|$  is not significantly larger than $k_BT$.}
\label{fig1}
\end{figure}

\begin{figure}[h]
\epsfxsize= 8 cm
\centerline{\epsfbox{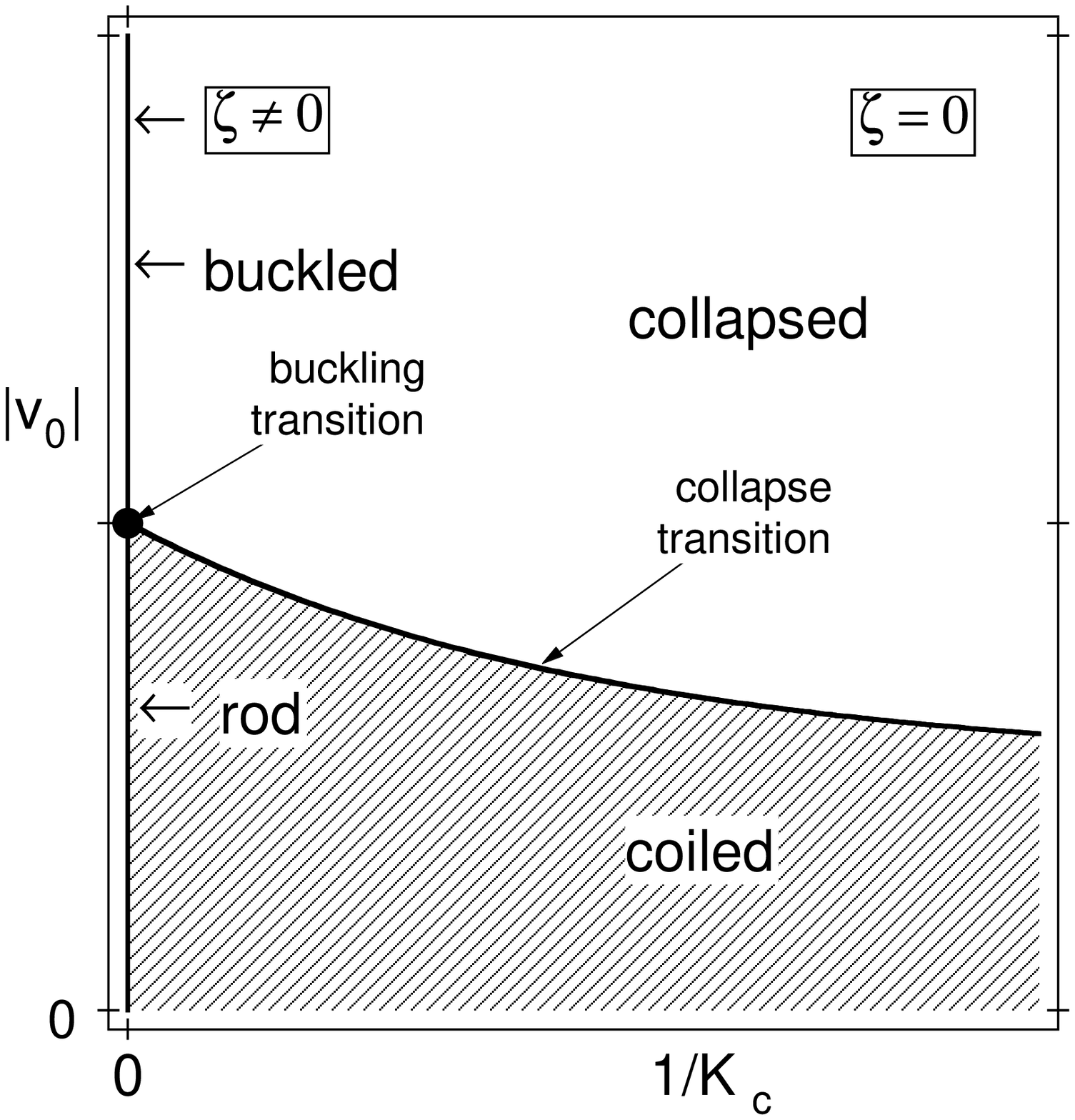}}
\caption{The phase diagram for a semi-flexible polymer whose
bending rigidity is $K_c$, and whose monomers interact via
an attractive potential of strength $|v_0|$. Buckling of a
rod may take place for some $|v_0|$ when the bending rigidity
is effectively infinite. Collapse of a random coil may take
place for some $|v_0(K_c)|$ when $K_c$ is finite.}
\label{fig2}
\end{figure}

The outcome of numerical and qualitative analysis is the following:
If one fixes $\kappa$ and ${\ell}_p$, and determines numerically how
${\ell}_{p}^{(R)}/{\ell}_p$ (or $\delta K_c/K_c$)
depends on $|v_0|$ one finds results, of which the curve shown
in  Fig.~\ref{fig1}:

The curve shows that for large values of ${\ell}_{p}^{(R)}/{\ell}_p$
(small values of $|\delta K_{c}|$) and small values of $v_0$, a
solution exists for which ${\ell}_{p}^{(R)}/{\ell}_p$ will decrease
$(|\delta K_{C}(v_0)/K_c|$ will increase) with $|v_0|$, as it should.
For values of  ${\ell}_{p}^{(R)}/{\ell}_p$ large compared
with $\kappa{\ell}_p$,
$|({\ell}_{p}^{(R)}-{\ell}_p)/{\ell}_p|=|\delta K_c/K_c|$ is
found to vary linearly with $|v_0|$. This same
result is found by a steepest descent analysis, and agrees with
the $T=0$ result displayed in Eq.~(\ref{dkap1}).  When $|v_0|$ is
increased
further, the decrease in ${\ell}_{p}^{(R)}/{\ell}_p$ (increase in
$|\delta K_{c}/K_c|$) speeds up. Eventually, for some $|v_{0c}|$,
the rate of change of ${\ell}_{p}^{(R)}/{\ell}_p$ and of
$|\delta K_{c}/K_c|$ is ``predicted'' to become infinitely
fast. It is worth noting that $|v_{0c}|$  is not significantly
larger than $k_BT$. Thus, whenever the monomer-monomer interaction
grows larger than the energy scale set by the thermal energy,
the chain will collapse.
If one then increases $v_{0}$ beyond $v_{0c}$, one finds no
solution for $\delta K_{c}/K_c$ which satisfies the demand that
$|\delta K_c/K_c|$
 increases as $|v_0|$. This result is in good agreement
with the results of a formal steepest descent analysis which, for
$\kappa{\ell}_p\rightarrow 0$, predicts that
\begin{eqnarray}
\delta K_{c}Ä = v_{0}\frac{\lambda_{s}^{6}}{{(K_{c}^{(R)})}^5}
\times{\cal{O}}(1)~,\end{eqnarray}
with a non-trivial relation between $\delta K_c$ and $v_0$ and the
screening length $\lambda_s=1/\kappa$.
Physically, the latter result is not acceptable and it must be
seen as an indication that the theory breaks down. In fact, we
believe that the solution $\zeta=0$ with correlations well described
by a random walk model will now have to be replaced by a solution
characterizing the collapsed state. If our analysis is correct, we can
therefore conclude that if we investigate the polymer at finite
temperatures (finite ``bare'' bending rigidity), we will find that not
only will it wish to bend, if the
interactions between monomers are strong enough, it will also prefer
to collapse,  in order to
overcome the entropic penalty. We illustrate these predictions of the
conformational behavior in the phase diagram, Fig.~\ref{fig2}.
The interesting feature in this phase diagram is the line of collapse
transitions which exist for finite values of $K_c$ and terminates
at the buckling instability point at $K_c=\infty$.
Based on our formal steepest descent calculation, we predict that for
large enough $1/K_c$, $v_c(1/K_c) \sim {(1/K_c)}^{-1/6}$.

\newpage
\section{Discussion}

In the above analysis we explored the connection between buckling of
self-interacting elastic rods
and polymer collapse because of
attractive segment-segment interactions.

This buckling transition is effectively the same as an Euler instability
under externally imposed compression forces \cite{LL}. We derived an
elastic equilibrium equation whose solutions determine the
state of the elastic rod in the presence of attractive segment-segment
interactions. In effect, the buckled state corresponds to the bound
state solutions of a Schr\" odinger-like equation to which the elastic
equilibrium equation is closely related. Except for extremely simple
interaction potentials this equation can of course not be solved
analytically. Nevertheless we find, that the WKB solution quite
accurately describes the qualitative as well as some of the
quantitative aspects of the numerical solution, especially with short
range potentials.

Qualitatively, the introduction of thermal fluctuations at the
harmonic level, does not change the  picture of the
buckling transition. It nevertheless points to the conclusion that
conformational fluctuations will renormalize the value of the
persistence length. This effect is quite well known, if not
understood in all its details, in the case of repulsive potentials
\cite{Odijk}, \cite{Skolnic&Fixman}, \cite{Joanny& Barrat} where the interactions tend
to stiffen up the chain. The attractive potential, not surprisingly,
acts in the reverse direction, thus diminishing the persistence
length. The harmonic approximation, valid strictly only in the limit
of small fluctuations, makes the buckling transition, where
fluctuations may become prohibitively large, difficult to analyze in
quantitative terms. We nevertheless argue that it is still there, but
displaced towards a different point in the parameter space. This
displacement is predicted to be linear in $(\beta K_c)^{-1}$.

For unconstrained fluctuations it is difficult to put forth a
comprehensive theory. We use the systematic $1/d$-expansion, which has
previously been applied in studies of higher dimensional
self-interacting manifolds \cite{LeDoussal}, \cite{Palmeri&Guitter} as a
vehicle to build a more general theory of the self-attracting polymer
chain. On the level of approximation provided by the systematic $1/d$
expansion it appears that buckling in the strict sense of this word is
preserved only at $T \rightarrow 0$ or, equivalently, for infinitely
stiff chains, $K_{c} \rightarrow \infty$. At any finite temperature,
or finite persistence length, the buckling transition is turned into a
collapse of the same type as already extensively investigated in the
case of self-attracting ideally flexible polymers \cite{shura}.

This scenario of course depends on the level of approximation
provided by the $1/d$-expansion in the $d \rightarrow \infty$
limit. Usually the variational approach, being non-perturbative, does
not fare  badly; we have confidence that the salient features of
the phase diagram for the self-interacting stiff polymer chain are not
far off from the picture put fourth here. The weakest link in our
story would appear to be the {\sl ansatz} ${\bf{r}}_{0}Ä=\zeta s
{\bf{e}}$. It obviously can not describe the more realistic toroidal
shapes of the {\sl e.g.} DNA aggregates. But it should certainly work
fine as long as we are not interested in the detailed structure of the
collapsed phase but only in the phase boundary.

At present detailed predictions for experiments are unrealistic. At
least the orientation dependent part of the interaction should be
included in order to describe the nematic nature \cite{grokho} of the
condensed state. One thing however we consider to be a robust result
of our calculations: counterion correlation attractions deminish the
persistence length. The opposite effect of the stiffening of the chain
with repulsive intersegment interactions is of course well known,
though perhaps less well understood \cite{Odijk},  
\cite{Skolnic&Fixman}, \cite{Joanny& Barrat}. The effect alluded to here is not just the OSF
formula with the sign reversed. It has a completely different
screening length and magnitude dependence than the OSF result.

The linearized version of this effect is embodied in
Eq.~(\ref{attrenorm}). Recent experiments on stretched DNA in the
presence of variable amount of $\rm Co(NH_3)_6^{3+}$ \cite{ruzvic}
clearly show that the effective persistence length gets smaller the
higher the concentratiuon of the condensing agent, that without doubt
confers some correlation attraction to the intersegment interaction
potential. In these experiments the concentration of the condensing
agent is too small to induce a full blown collapse of DNA, but still,
the incipient effects are seen in the smaller effective peristence
length. Qualitatively this is exactly what one expects from our
theory.

Also the present form of the theory seems to be well suited to
describe the effects of the correlation attractions on the elastic
extension of the chain. The interplay between collapse and stretching
seems to be well within the reach of the present formalism and will be
pursued in all the details in the future \cite{RudiPer}.

\section{Acknowledgements:}
We would like to thank Murugappan Muthukumar, Tanniemola Liverpool,
Ramin Golestanian, Mehran Kardar for stimulating discussions. This
research was supported in part by the National Science Foundation
under Grant No.PHY94-07194.

\vfill
\eject

\vfill
\eject

\section{Figure Captions:}

{\bf Figure 1:}  Schematic representation of buckling. In a buckled chain the 
effective separation between segments becomes smaller which lowers 
the free energy of the chain if the interactions between segments are 
attractive. This decrease in the free energy works against  the 
increased bending energy thus leading to an instability.

\smallskip

{\bf Figure 2:} Critical strength of the interaction potential, in terms of
$V_{cr}~L^2$, as a function of the length of the rod, $L$, at
different values of the screening parameter $r_0$ for the three models of attractive
potentials: Finite potential well, Eq.~(\ref{potential}), exponential
potential, Eq.~(\ref{exponential}), and the general correlation
potential, Eq.~(\ref{correlation}). The dependencies have been rescaled in such a
way that at large $L$ they coincide.

\smallskip

{\bf Figure 3:} The solution of Eq.~(\ref{rigid}), the relation between
the strength of segmental attraction and the change in apparent
bending modulus, for $\kappa {\ell}_p=0.1$. Observe that
${\ell}_{p}^{(R)}/{\ell}_p$ depends linearly on $|v_0|$ for small
values of $|v_0|$. There exists a $|v_{0c}|$ where there is no longer a physically acceptable solution
for ${\ell}_{p}^{(R)}/{\ell}_p (|v_0|)$, implying a loss of stability 
of the coiled configuration of the polyelectrolyte chain. It is worth 
observing that $|v_{0c}|$  is not significantly larger than $k_BT$.

\smallskip

{\bf Figure 4:} The phase diagram for a semi-flexible polymer whose
bending rigidity is $K_c$, and whose monomers interact via
an attractive potential of strength $|v_0|$. Buckling of a
rod may take place for some $|v_0|$ when the bending rigidity
is effectively infinite. Collapse of a random coil may take
place for some $|v_0(K_c)|$ when $K_c$ is finite.

\vfill
\eject

\section{Figures:}

\begin{thebibliography}{55}
\bibitem{NOTE} The formalism developed here has the following pleasing
         feature. If we choose a Debye-H\"{u}ckel interaction between
         monomers, and assume $\zeta=1$ ($T=0$), we determine an
electrostactic
         contribution to the rigidity, in quantitative agreement with the
         celebrated results obtained by Odijk \cite{Odijk};  Skolnick \& Fixman
         \cite{Skolnic&Fixman}; and Joanny \& Barrat \cite{Joanny& Barrat}.
\end{thebibliography}
\end{document}